\input{aipcheck.tex}

   \def\selectedoptions{final}

\documentclass[
   \selectedoptions
  ]
  {aipproc}

\newcommand\Eiso{E$_\mathrm{iso}$}

\newcommand\Epeak{E$_\mathrm{peak}$}
\newcommand\epeak{e$_\mathrm{peak}$}
\newcommand\Ngam{N$_\gamma$}
\newcommand\ngam{n$_\gamma$}
\newcommand\Xpar{\Ngam /(\Epeak *$\sqrt{T_{90}}$)}
\newcommand\xpar{\ngam /(\epeak *$\sqrt{t_{90}}$)}

\def\selectedlayoutstyle{6x9}
\layoutstyle\selectedlayoutstyle
\SetInternalRegister\hbadness{8000} 

\newcommand\doingARLO[2][]{%
  \ifx\mmref\undefined #1\else #2\fi
}

\begin{document}

\title 
      []
      {Observation and implications of \\
       the \Epeak - \Eiso\ correlation in Gamma-Ray Bursts}


\author{J-L. Atteia}{
  address={Laboratoire d'Astrophysique, Observatoire Midi-Pyr\'en\'ees,
  Toulouse, France}
}
\author{G. R. Ricker}{
  address={Center for Space Research,
  Massachussetts Institute of Technology,
  Cambridge, MA 02139, USA}
}
\author{D. Q. Lamb}{
  address={Department of Astronomy \& Astrophysics, University of Chicago,
  Chicago, IL 60637, USA}
}
\author{T. Sakamoto}{
  address={Tokyo Institute of Technology,
  2-12-1 Ookayama, Meguro-ku, Tokyo 152-8551, Japan},
  altaddress={RIKEN (The Institute of Physical and Chemical Research),
  Saitama 351-0198, Japan}
}
\author{C. Graziani}{
  address={Department of Astronomy \& Astrophysics, University of Chicago,
  Chicago, IL 60637, USA}
}
\author{T. Donaghy}{
  address={Department of Astronomy \& Astrophysics, University of Chicago,
  Chicago, IL 60637, USA}
}
\author{C. Barraud}{
  address={Laboratoire d'Astrophysique, Observatoire Midi-Pyr\'en\'ees,
  Toulouse, France}
}
\author{The HETE-2 Science Team}
{address={An international collaboration of institutions including MIT, LANL,
U. Chicago, U.C. Berkeley, U.C. Santa Cruz (USA), CESR, CNES, Sup'Aero (France),
RIKEN, NASDA (Japan), IASF/CNR (Italy), INPE (Brazil), TIFR (India)}}

\copyrightyear  {2004}

\begin{abstract}
The availability of a few dozen GRB redshifts now allows studies 
of the intrinsic properties of these high energy transients. 
Amati et al. recently discovered a correlation between  \Epeak, 
the intrinsic peak energy of the $\nu f \nu$ spectrum, and \Eiso, 
the isotropic equivalent energy radiated by the source. 
Lamb et al. have shown that HETE-2 data confirm and extend this correlation. 
We discuss here one of the consequences of this correlation: 
the existence of a 'spectral standard candle', which can be used 
to construct a simple redshift indicator for GRBs.
\end{abstract}

\date{\today}

\maketitle

\section{The \Epeak $-$ \Eiso\ relation for Gamma-Ray Bursts}

\begin{figure}[h]
\label{lambfig}
\includegraphics[width=0.5\columnwidth]{atteia_fig0.ps} 
\includegraphics[width=0.5\columnwidth]{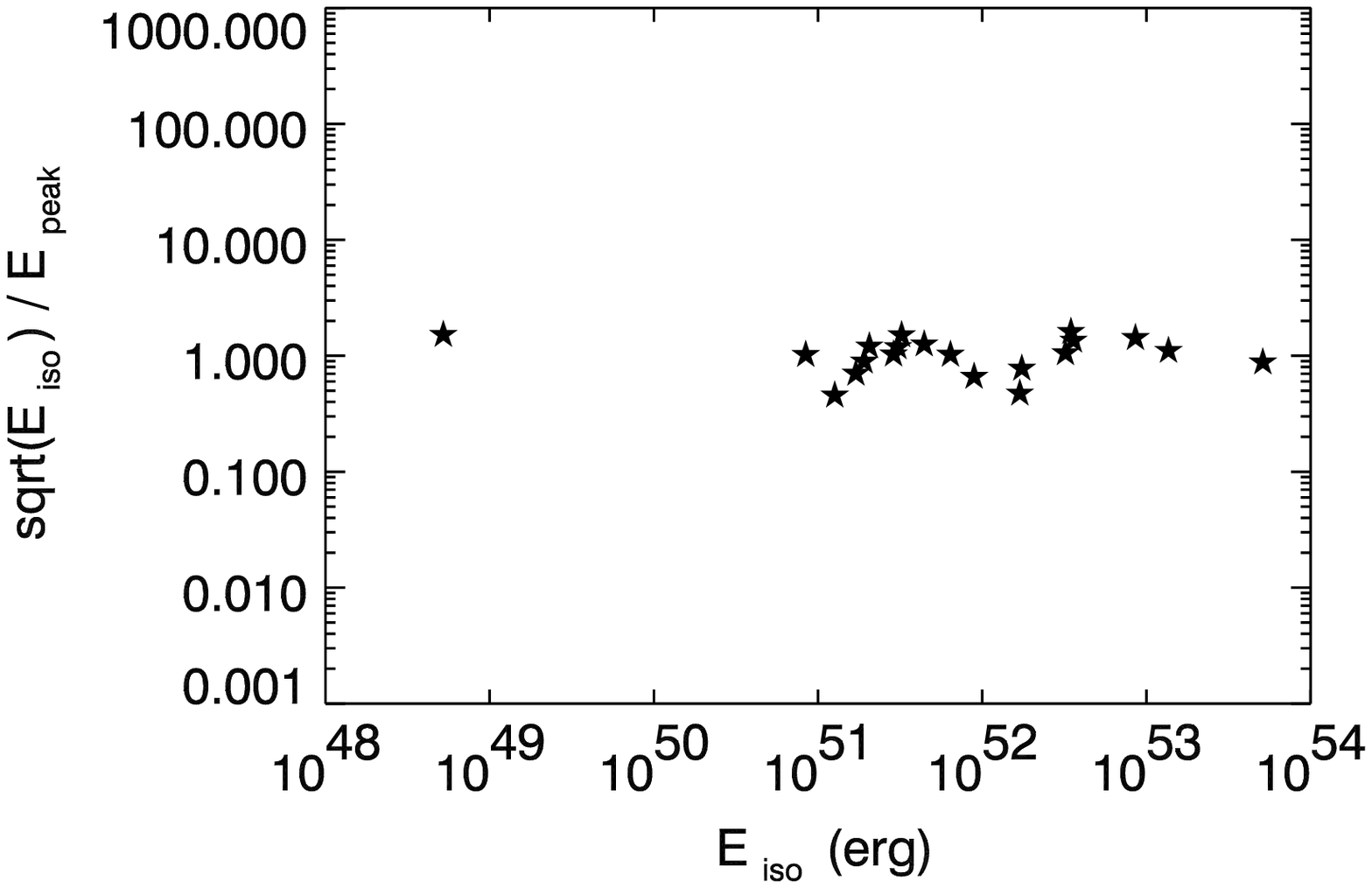} 
\caption{{\it Left Panel.} The \Epeak - \Eiso\ correlation measured at the end of 2003 with 21
GRBs detected by BeppoSAX (Amati et al. 2002), HETE-2 (Sakamoto et al. 2003, Lamb et al. 2003), 
and the IPN (Andersen et al. 2000). Note the extent of the correlation in \Eiso . 
{\it Right Panel.} Illustration of the fact that the ratio $\sqrt{E_\mathrm{iso}}$ / \Epeak\ is close 
to a standard candle. This ratio appears almost constant over 4-5 orders of magnitude
in \Eiso . The ratio $\sqrt{E_\mathrm{iso}}$ / \Epeak\ is plotted here for 20 GRBs with known
redshift detected with BeppoSAX, HETE-2, and the IPN.
}
\end{figure}

The growing sample of GRBs with spectroscopic redshifts allows the measure
of some {\it intrinsic} properties of these explosions, 
like the energy radiated at various wavelengths or the energy at which
most of the power is emitted.
In 2002, Amati et al. performed a systematic analysis of 12 GRBs 
with known redshifts in order to derive their 
intrinsic spectral parameters (the spectral parameters at the source).
In their paper, they report a strong correlation between  \Epeak , 
the intrinsic peak energy of the $\nu f \nu$ spectrum, and \Eiso ,
the isotropic equivalent energy radiated by the source.
For many years this correlation was suspected because it is the best way 
to explain the well known Hardness-Intensity correlation observed in GRBs 
(e.g. Mallozzi et al. 1995, Dezalay et al. 1997, Lloyd et al. 2000, Lloyd-Ronning \& Ramirez-Ruiz 2002,
and ref. therein).
However the lack of distance measurements kept the work on the hardness-luminosity correlation qualitative.
Recently Lamb et al. (2003) pointed out that HETE observations not only confirm this 
correlation, but also suggest its extension to the population of X-Ray Flashes (Fig. \ref{lambfig}a).
According to (Fig. \ref{lambfig}a), the \Epeak - \Eiso\ correlation 
can be approximated by the following relation:
$\mathrm{E_{peak} / (100\ keV) = \sqrt{E_{iso} / (10^{52} erg)}}$.
The origin of this correlation (which is reminiscent of the Temperature-Luminosity
correlation for clusters of galaxies) is not discussed here. 
The aim of this paper is to discuss one of its consequences: 
the existence of a 'spectral standard candle', which can be used to 
construct a redshift indicator for GRBs.

\section{Building a redshift indicator for GRBs}

\begin{figure}
\label{zindic}
\begin{tabular}{cc}
\includegraphics[width=0.5\columnwidth]{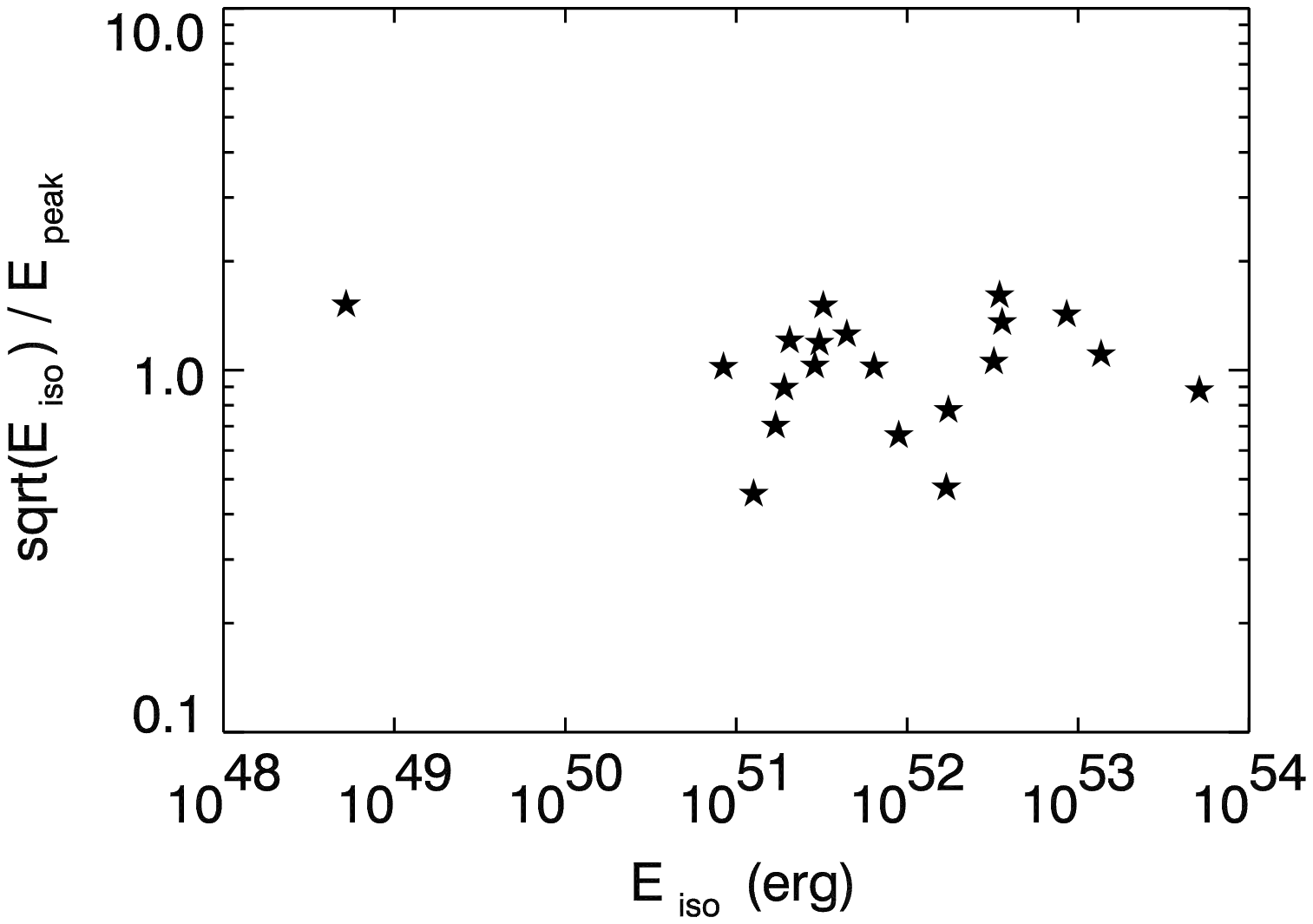} & \includegraphics[width=0.5\columnwidth]{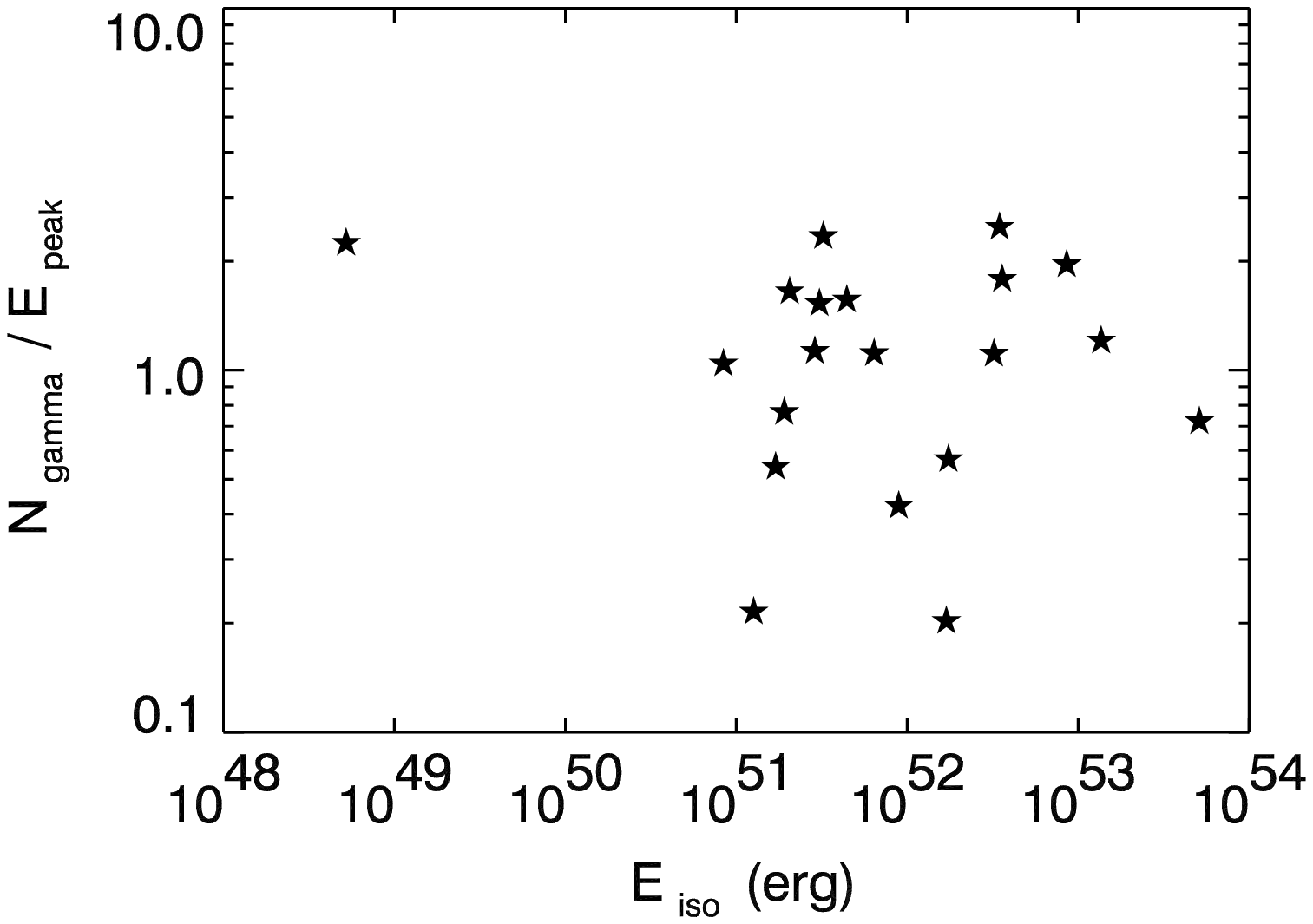} \\
\includegraphics[width=0.5\columnwidth]{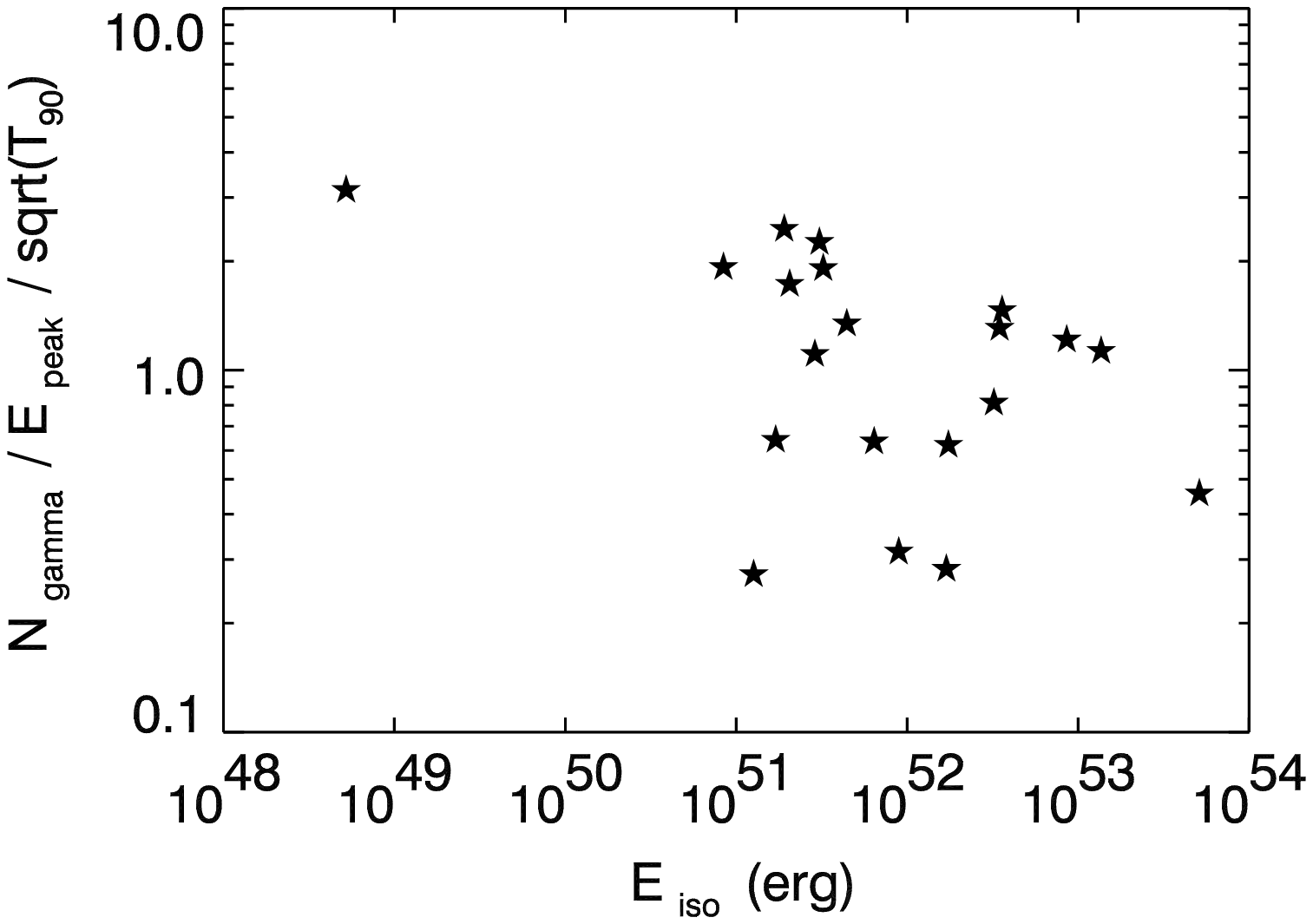} & \includegraphics[width=0.5\columnwidth]{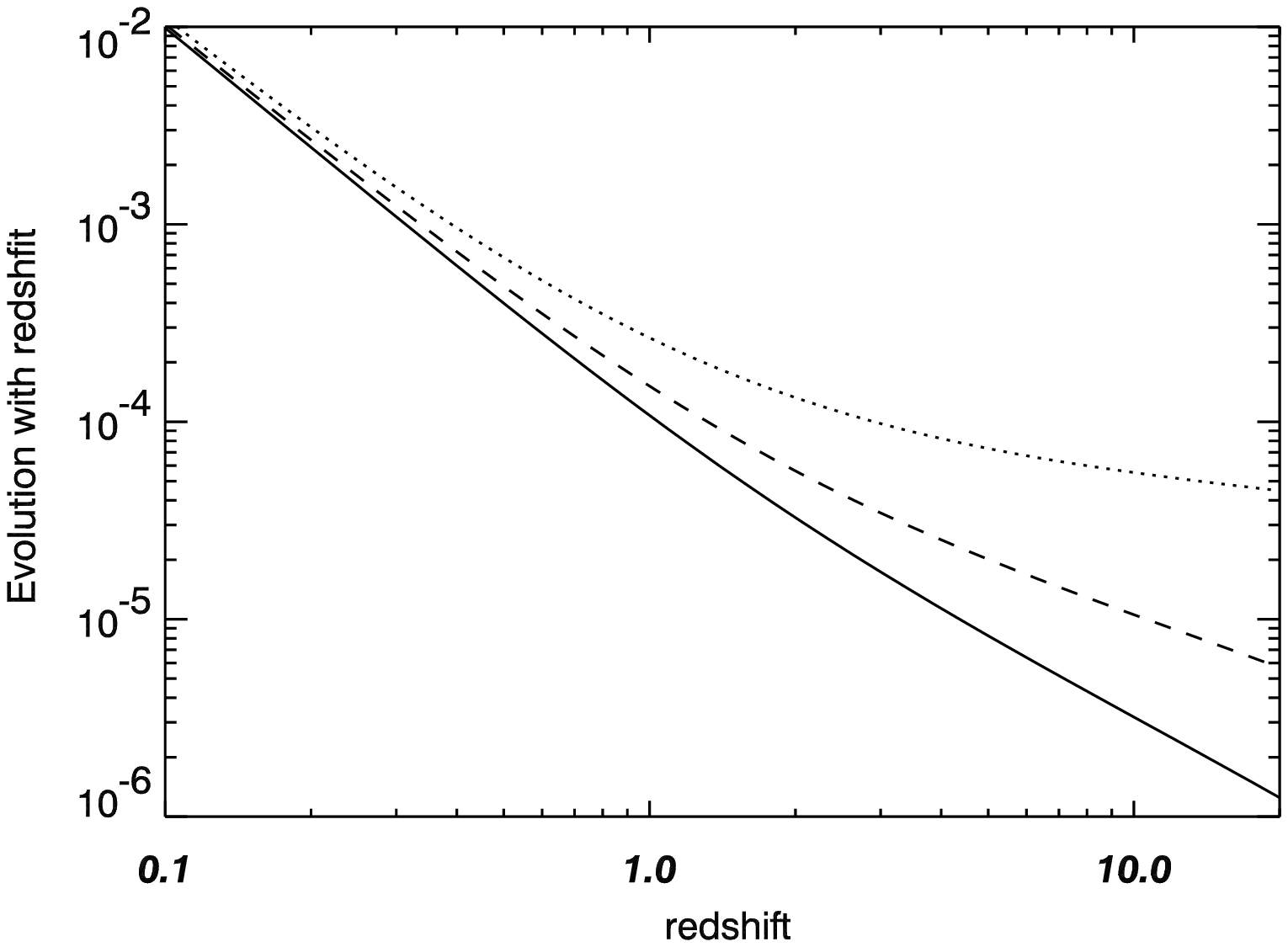} \\
\end{tabular}
\caption{Comparison of various redshift indicators. The lower right panel shows the 
theoretical dependence on redshift of $\sqrt{E_\mathrm{iso}}$ /\Epeak\ (dotted line), 
\Ngam /\Epeak\ (dashed line), and \Ngam /(\Epeak *$\sqrt{T_{90}}$) (solid line), 
where \Ngam\ is the number of photons emitted by the GRB between (\Epeak /100) and (\Epeak /2),
and T$_{90}$ is the duration of the burst .
The other three panels show the intrinsic dispersion of these ratios.}
\end{figure}

The good correlation between \Epeak\ and \Eiso\ suggests that the
ratio $\sqrt{E_\mathrm{iso}}$ / \Epeak\ is close to a standard candle.
This is illustrated in Fig. \ref{lambfig}b which shows this ratio as a function 
of \Eiso\ for 20 gamma-ray bursts with known redshifts.
Assuming that  $\sqrt{E_\mathrm{iso}}$ / \Epeak\ is constant at the source,
it is easy to compute its evolution with redshift.
This is the dotted curve in the lower right panel of Fig. \ref{zindic}.
This curve shows that unfortunately the {\it observed} ratio has a very small 
dependence on redshift beyond z=1.
This illustrates the fact that when one wants to find a redshift indicator,
it must find a quantity which has not only a small intrinsic dispersion, but also
a strong dependence on redshift over a large range of redshifts.
We performed an empirical search for such a quantity, starting
from the fact that $\sqrt{E_\mathrm{iso}}$ / \Epeak\ is close to a
standard candle. This work led us to conclude that the quantity X$_0$ = \Xpar\
has the right properties for a redshift indicator (we do not
claim however that it is the best redshift indicator which can be constructed from 
gamma-ray data only).
Fig. \ref{zindic} shows the intrinsic dispersion of X$_0$ (lower left panel), 
and its dependence on redshift (solid curve in the lower right panel).
In the definition of X$_0$, \Epeak\ is the peak energy of the $\nu f \nu$ spectrum,
\Ngam\ is the number of photons emitted by the GRB between (\Epeak /100) and (\Epeak /2),
and T$_{90}$ is the duration of the burst (see Atteia 2003 for additional precisions
on X$_0$). All these parameters are measured at the source.

\begin{figure}
\label{pzvsz}
  \includegraphics[width=0.5\columnwidth]{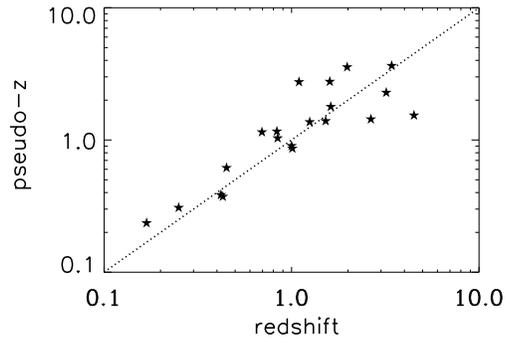}
  \caption{Pseudo-redshifts of 20 GRBs detected by BeppoSAX and HETE,
compared with their spectroscopic redshifts.}
\end{figure}

Using X$_0$ as a redshift indicator, we can compute pseudo-redshifts 
by assuming that X$_0$ at the source is constant and 
that the observed value X=\xpar\ (the lower case letters indicating that
the parameters are now measured in the observer's framework) 
differs from X$_0$ only for the effect of the redshift.
The validity of these pseudo-redshifts can be assessed from Fig. \ref{pzvsz} which compares pseudo-redshifts
and spectroscopic redshifts of 20 GRBs detected and localized by BeppoSAX, HETE-2, and the IPN.
Possible applications of these pseudo-redshifts are discussed in Atteia (2003).

\section{Pseudo-redshifts of HETE-2 GRBs}

Table \ref{tab1} presents the pseudo-redshifts of 42 long GRBs detected by HETE-2, 
with comments on their spectral properties, and on the detection of an afterglow.
Pseudo-redshifts range from 0.20 (GRB 030824) to 14.0 (GRB 031026). 
Nine of the GRBs in Table \ref{tab1} have spectroscopic redshifts (in the range 0.25 to 3.2). 
The pseudo-redshifts of these bursts are all within a factor of two 
of the spectroscopic redshifts, which leds us to conclude that pseudo-redshifts
provide a robust redshift indicator in the range z=0.2-3.
Table \ref{tab1} contains three GRBs (in boldface)
which have pseudo-redshifts larger than 5, 
suggesting that they could be GRBs at high redshifts. 
Unfortunately the spectroscopic redshifts of these three GRBs have not been measured,
leaving open the issue of whether pseudo-redshifts are
reliable at large redshifts.
If pseudo-redshifts are reliable beyond z=5-6,
they may become a useful tool 
to quickly identify high-z GRBs, and trigger the 
follow-up actions which are appropriate for these bursts 
(e.g. X-ray and IR observations).

Finally, we would like to mention that after discussions
with the participants at the grb2003 Conference in Santa Fe, 
the HETE-2 Science Team and Operation Team now routinely provide 
the spectral parameters and the pseudo-redshifts of GRBs localized by HETE-2.
These parameters are made available to the community on a web page 
a few minutes only after the determination of the burst localization
(see GCNs 2421 and 2444). 

\begin{theacknowledgments}
The authors acknowledge the wonderful work of the HETE operation team.
\end{theacknowledgments}


\begin{table}[t]
\label{tab1}
\begin{tabular}{lllllllll}
\hline
\tablehead{1}{l}{b}{Name \\ }
  & \tablehead{1}{l}{b}{redshift\\ }
  & \tablehead{1}{l}{b}{pseudo- \\ redshift}
  & \tablehead{1}{l}{b}{Comment \\}
  & \tablehead{1}{r}{b}{\hspace{1cm} \\}
  & \tablehead{1}{l}{b}{Name \\}
  & \tablehead{1}{l}{b}{redshift \\}
  & \tablehead{1}{l}{b}{pseudo- \\ redshift}
  & \tablehead{1}{l}{b}{Comment \\ }   \\
\hline
grb001225       &      & 0.69       & Bright GRB             && grb021104       &      & 0.88       & X-Ray Flash         \\
grb010126       &      & 1.52       &                        && grb021211       & 1.01 & 0.86       & OA                  \\
grb010213       &      & 0.23       & X-Ray Flash            && grb030115       &      & 1.44       & X-Ray Rich          \\
grb010326       &      & 3.43       &                        && grb030226       & 1.98 & 3.56       & XA, OA              \\
{\bf grb010612} &      & {\bf 9.50} &                        && grb030324       &      & 3.93       & dark, X-Ray Rich    \\
grb010613       &      & 0.70       &                        && grb030328       & 1.52 & 1.39       & OA, XA              \\
grb010629       &      & 0.57       & X-Ray Rich             && grb030329       & 0.17 & 0.24       & Bright, OA, XA, RA, SN 2003dh, X-Ray Rich \\
grb010921       & 0.45 & 0.62       & OA, RA                 && grb030418       &      & 1.10       & X-Ray Flash         \\
grb010928       &      & 3.22       &                        && grb030429       & 2.65 & 1.44       & OA                  \\
grb020124       & 3.20 & 2.28       & OA                     && grb030519       &      & 2.53       &                     \\
grb020127       &      & 2.67       & XA, RA, X-Ray Rich     && grb030528       &      & 0.36       & IRA, XA             \\
{\bf grb020305} &      & {\bf 5.88} & {\bf OA}               && grb030723       &      & 0.59       & OA, XA, SN, X-Ray Flash \\
grb020317       &      & 1.86       & X-Ray Flash            && grb030725       &      & 1.21       & OA                  \\
grb020331       &      & 2.90       & OA                     && grb030821       &      & 2.36       & X-Ray Rich          \\
grb020418       &      & 1.92       &                        && grb030823       &      & 0.64       & X-Ray Flash         \\
grb020801       &      & 0.95       & X-Ray Rich             && grb030824       &      & 0.20       & X-Ray Flash         \\
grb020812       &      & 3.03       &                        && {\bf grb031026} &      & {\bf 14.0} &                     \\
grb020813       & 1.25 & 1.37       & OA, XA, RA             && grb031109a      &      & 1.29       &                     \\
grb020819       &      & 1.52       & RA, X-Ray Rich         && grb031109b      &      & 1.39       & X-Ray Flash         \\
grb020903       & 0.25 & 0.31       & OA, RA                 && grb031111a      &      & 4.20       &                     \\
grb021016       &      & 1.45       &                        && grb031111b      &      & 0.56       & X-Ray Rich          \\
\hline
\end{tabular}
\caption{Pseudo-redshifts of 42 long GRBs detected by HETE. The pseudo-redshifts of 
a few HETE GRBs could not be calculated because their \epeak\ 
is outside the energy range of HETE (e.g. GRB 021004). 
The 'Comment' column indicates the spectral hardness of the burst, and 
the detection of an afterglow when appropriate. OA, XA, RA, and IRA respectively stand for
Optical Afterglow, X-ray Afterglow, Radio Afterglow, and Infra-Red Afterglow.
Three GRBs with pseudo-redshifts greater than 5 are indicated in bold.}
\end{table}


\begin{thebibliography}{}
\bibitem{amat02} Amati, L., Frontera, F., Tavani, M., ~et al. \ 2002, A\&A, 364, L54
\bibitem{ande00} Andersen, M.L., Hjorth, J., Pedersen, H.~et al. \ 2000, A\&A, 390, 81
\bibitem{atte03} Atteia, J-L. \ 2003, A\&A, 407, L1
\bibitem{barr02} Barraud, C., Olive, J-F., Lestrade, J.P., et al. 2002, A\&A, 400, 1021
\bibitem{deza97} Dezalay, J-P., Atteia, J-L., Barat, C. et al. 1997, ApJ, 490, L17
\bibitem{lamb03} Lamb, D.Q., ~et al. \ 2003, ApJ, submitted
\bibitem{lloy00} Lloyd, N.M., Petrossian, \& V., Mallozzi, R.S. \ 2000, Apj, 534, 227
\bibitem{lloy02} Lloyd-Ronning, N.M., \& Ramirez-Ruiz, E. \ 2002, Apj, 576, 101
\bibitem{mall95} Mallozzi, R.S., Paciesas, W.S., Pendleton, G.N., et al. \ 1995, Apj, 454, 597
\bibitem{saka03} Sakamoto, T.,  et al. \ 2003, Apj, in press
\end{thebibliography}
\end{document}